\documentclass[11pt]{article}
\newcommand{\la}[1]{\label{#1}}
\newcommand{\be}{\begin{equation}}
\newcommand{\ee}{\end{equation}}
\newcommand{\ba}{\begin{eqnarray}}
\newcommand{\ea}{\end{eqnarray}}
\newcommand{\bi}{\begin{itemize}}
\newcommand{\ei}{\end{itemize}}
\newcommand{\rmi}[1]{{\mbox{\scriptsize #1}}}
\newcommand{\nr}[1]{(\ref{#1})}
\newcommand{\tr}{{\rm Tr\,}}

\newcommand{\nn}{\nonumber \\}
\newcommand{\fr}[2]{{\frac{#1}{#2}}}

\renewcommand{\vec}[1]{{\bf #1}}

\newcommand{\RR}{{\rm I\kern -.2em  R}}
\newcommand{\eq}{Eq.~}
\newcommand{\eqs}{Eqs.~}

\def\lsi{\raise0.3ex\hbox{$<$\kern-0.75em\raise-1.1ex\hbox{$\sim$}}}
\def\gsi{\raise0.3ex\hbox{$>$\kern-0.75em\raise-1.1ex\hbox{$\sim$}}}

\makeatletter
\renewcommand\section{\@startsection {section}{1}{\z@}%
                                   {-5.5ex \@plus -1ex \@minus -.2ex}
                                   {2.3ex \@plus.2ex}%
                                   {\normalfont\large\bfseries}}
\renewcommand\subsection{\@startsection{subsection}{2}{\z@}%
                                     {-3.25ex\@plus -1ex \@minus -.2ex}%
                                     {1.5ex \@plus .2ex}%
                                     {\normalfont\normalsize\bfseries}}
\renewcommand\thesection {\@arabic\c@section}
\renewcommand\thesubsection   {\thesection.\@arabic\c@subsection}
\renewcommand{\@seccntformat}[1]{%
\csname the#1\endcsname.\hspace{1.0em}}
\makeatother

\begin{document}

\begin{titlepage}
\begin{flushright}
CERN-TH/2001-309\\
hep-ph/0111113\\
\end{flushright}
\begin{centering}
\vfill

\mbox{\bf A REMARK ON NON-ABELIAN CLASSICAL KINETIC THEORY}

\vspace{0.8cm}

Mikko Laine\footnote{mikko.laine@cern.ch} and
Cristina Manuel\footnote{cristina.manuel@cern.ch}

\vspace{0.3cm}
{\em
Theory Division, CERN, CH-1211 Geneva 23,
Switzerland\\}

\vspace*{0.8cm}

\end{centering}

\noindent
It is known that non-Abelian classical kinetic theory reproduces
the Hard Thermal/Dense Loop (HTL/HDL) effective action of QCD,
obtained after integrating out the hardest momentum scales from
the system, as well as the first higher dimensional operator
beyond the HTL/HDL level. We discuss here its applicability at
still higher orders, by comparing the exact classical effective
action obtained in the static limit, with the 1-loop quantum effective
potential. We remark that while correct types of operators arise,
the classical colour algebra reproduces correctly the prefactor 
of the 4-point function $\tr A_0^4$ only for matter in asymptotically 
high dimensional colour representations. 
\vfill
\noindent


\vspace*{1cm}

\noindent
CERN-TH/2001-309\\
November 2001

\vfill

\end{titlepage}


\section{Introduction}

Most observables of QCD at a finite temperature $T$
and chemical potential $\mu$ are not computable in
perturbation theory beyond a certain
order, due to severe infrared problems~\cite{linde}.
What can be done systematically however
is the construction of various effective theories, obtained by
integrating out only the hardest scales from the system.
For static observables this leads to the concept of
a dimensionally reduced effective theory~\cite{dr},
whose construction and non-perturbative
properties have been studied in great detail
(for reviews, see~\cite{drrevs}).

For non-static observables, on the other hand,
the relevant effective description is the Hard Thermal
and/or Dense Loop (HTL/HDL) theory~\cite{rdp,bp} or,
equivalently, its local reformulation~\cite{bi,kelly}
as classical kinetic theory~\cite{Wong,uh1}.
These constructions and their non-perturbative properties have however
so far not been studied to a similar beyond-the-leading-order
level as those of the dimensionally reduced theory
(for a review of the current status, see~\cite{db}).

Very recently, there was a new positive indication on the
effectiveness of the kinetic description: it was noted that it
reproduces also the first higher order operator beyond the HTL/HDL
level~\cite{bl}, representing an effective (charge conjugation
violating) three-gluon interaction. In this brief note, we wish to
report on some properties of the classical kinetic theory at still
higher orders. We point out that in this case agreement with
quantum field theory is only obtained for matter in high
dimensional colour representations.

\section{Formulation of classical kinetic theory}

We start by reviewing briefly the formulation of classical kinetic
theory, used to describe how
``hard'' particles (quarks and gluons with momenta $\sim {\rm
max}(T,\mu)$), behave in the background of a ``soft''
($\sim {\rm max}(gT, g\mu)$, where $g$ is the gauge
coupling) gauge field configuration $A^a_\mu$.

The starting point is to consider the hard modes as
classical point particles carrying a
colour charge $Q^a$, with dynamics governed by the Wong
equations~\cite{Wong}. When the effect of collisions is
neglected, the 1-particle distribution function
obeys the Boltzmann equation~\cite{uh1}
\be
 \label{NA-f}
 p^\mu\left(\frac{\partial}{\partial x^\mu}
 + g f^{abc} Q^a A^{b}_\mu \frac{\partial}{\partial Q^c}
 - g Q^aF^{a}_{\mu\nu}\frac{\partial}{\partial p_\nu}\right)
 f(x,p,Q)=0 .
 \la{boltzmann}
\ee
Our sign conventions correspond to QCD
with a covariant derivative in the fundamental
representation $D_\mu = \partial_\mu - i g T^a A^a_\mu$,
and in the adjoint
${\cal D}_\mu^{ab} = \delta^{ab} \partial_\mu + g f^{acb} A^c_\mu$;
the field strength is $F_{\mu\nu} = (i/g)[D_\mu,D_\nu]$.
We take $N_f$
flavours of massless quarks and antiquarks, each carrying
two helicities. For antiquarks one
should replace $Q \rightarrow -Q$ in \eq\nr{boltzmann}, and
for gluons one should take the adjoint representation,
$Q \to Q_\rmi{adjoint} \equiv \tilde Q$.

For quarks/antiquarks the boundary condition for the solutions of
\eq\nr{boltzmann} is that for a vanishing gauge field background
(say, at infinity), 
\be
 f(x,p,Q) \to \delta_+(p^2)\, n_{f} (\omega_p \mp \mu),
 \quad
 \delta_+(p^2) \equiv
 2 \theta(p_0)\, (2\pi)\, \delta(p^2)\, ,
\la{deltap}
\la{bc}
\ee
where $n_f$ is the Fermi-Dirac distribution
function, and $\omega_p = |\vec{p}|$.
For gluons $f(x,p,Q) \to \delta_+(p^2)\, n_{b} (\omega_p)$,
where $n_b$ is the Bose-Einstein distribution function.

The solution of \eq\nr{boltzmann} defines a current induced by the
coloured particles,
\be
 j_\nu^a(x) = g \sum_{\hbox{\tiny helicities}\atop\hbox{\tiny species}}
 \int_{p,Q} p_\nu Q^a f(x,p,Q), \la{jmu} 
\ee
with
\be
 \int_p
 = \int \frac{dp_0}{(2\pi)}\int_\vec{p}, \quad \int_\vec{p} = \int
 \frac{d^3p}{(2\pi)^3}, \quad \int_{Q} = \int dQ . 
\ee 
Here $dQ$ is
the colour measure, which contains delta functions fixing the
representation dependent Casimirs (see, e.g.,~\cite{lm});
we shall return to its properties presently.
The current in \eq\nr{jmu}, in turn, defines via
\be
 j^\nu_a = - \frac{\delta}{\delta {A_\nu^a}} \delta S_M, \la{SM}
\ee
an action $S_M  =  S_M^\rmi{local} + \delta S_M$, where
$S_M^\rmi{local} = -\int_x (1/2) \tr F_{\mu\nu}F^{\mu\nu}$.

There are actually two somewhat different formulations of the
kinetic theory, corresponding to the orders in which the integrals
$\int_p, \int_Q$ are to be carried out in the above. We shall
mostly carry out $\int_p$ first. However, if one takes the moments
$\int_Q\, (...), \int_Q Q^a\,(...)$ of \eq\nr{boltzmann}, imposing
by hand the QCD type relation (say, for the fundamental representation)
\be \label{rel-imp}
 \int_Q Q^a Q^b f(x,p,Q) = \frac{1}{2N_c}
 \delta^{ab} \bar f(x,p) + \fr12 d^{abc} f^c(x,p),
\ee
with $\bar f = \int_Q f,\quad f^a = \int_Q Q^a f$, then one gets a closed set
of equations for $\bar f, f^a$~\cite{uh1}: 
\ba 
& &
 p\cdot \partial \bar f  - g p^\mu F^a_{\mu\nu}
 \frac{\partial f^a}{\partial p_\nu} = 0, \la{jmua1} \\
& &
  (p\cdot {\cal D})^{ab} f^b
+ \frac{g}{2} d^{abc} p^\mu F^b_{\mu\nu}
  \frac{\partial f^c}{\partial p_\nu}
- \frac{g}{2N_c} p^\mu F^a_{\mu\nu}
  \frac{\partial \bar{f}}{\partial p_\nu} = 0. \la{jmua}
\ea
We refer to this as the second formulation. Since the colour
hierarchy is truncated by~\eq\nr{rel-imp}, the
first and second formulations only agree for the low order colour
moments.

\section{Exact solutions for special backgrounds}

Let us now recall that, for some special background
field configurations, one can find an ansatz leading
to an exact solution of \eq\nr{boltzmann}. Such solutions
have previously been discussed in the Abelian case~\cite{hakim,weert}
and, for static backgrounds,
in the non-Abelian~\cite{uh1} (see also~\cite{bft}).
However no comparison has been made with QCD, as far as we know.

We will search for a solution of \eq\nr{boltzmann} in a form that
only depends on the canonical momenta. Introducing the shorthands
\be
 A^a Q^a \equiv A, \quad A^a \tilde Q^a \equiv \tilde A,
\ee
we take for quarks (with a set of $\alpha$'s to be specified presently)
\be 
 f(x,p,Q) = \delta_+(p^2)\, F(p_\alpha + g A_\alpha). 
\la{ansatz} 
\ee 
Plugging into \eq\nr{boltzmann}, we
immediately find that if $\alpha$'s exist such that the condition 
$\partial_\alpha A^a_\nu = 0$ is satisfied for all $a, \nu$, then the ansatz
in~\eq\nr{ansatz} solves~\eq\nr{boltzmann}. Similar solutions hold
for antiquarks, by replacing $Q \rightarrow -Q$, and for gluons,
by replacing $Q \rightarrow \tilde Q$.

Apart from~\eq\nr{boltzmann}, the ansatz should
also satisfy the boundary condition in~\eq\nr{bc}.
Thus, in the static limit ($\partial_0 A^a_\nu =0$), we obtain (for quarks)
\be
 f(x,p,Q) = \delta_+(p^2)\,
 n_{f}(p_0 + g  A_0 -\mu). \la{stat}
\ee
An exact solution could also be obtained
in the homogeneous limit, $\partial_i A^a_\nu =0$,
by replacing $p_0 + g A_0$  in \eq\nr{stat}
by $|\vec{p} + g \vec{A}|$.

One may of course ask whether these exact solutions are the most
general ones, given the boundary conditions in \eq\nr{bc}. It is
at least easy to verify iteratively (without any ansatz) that they
do agree with the perturbative solution in $g Q^a$ of
\eq\nr{boltzmann} around $f^{(0)}$ defined by~\eq\nr{bc}: writing
$f= f^{(0)} + f^{(1)} + f^{(2)} + ...$, the $n^{\rm th}$ order
solution is obtained by solving 
\be 
 p^\mu {\hat D}_\mu f^{(n)} = {\hat L} f^{(n-1)} \ , 
\ee
where ${\hat D}_\mu = \partial_\mu + g
f^{abc} Q^a A^{b}_\mu \partial_Q^{c}$, ${\hat L} = g p^\mu Q^a
F^a_{\mu \nu} \partial^\nu_p$.

\section{Comparison with quantum field theory}

Let us now see what kind of an effective action, $S_M$, the
solutions found in the previous section lead to, and compare with quantum
field theory.

Plugging \eq\nr{stat} into \eq\nr{jmu} and taking into account the $N_f$
flavours and the two helicities,
we obtain, say, for the quark
and antiquark contribution to the gauge current,
\ba
 j^a_0 & = & 2 g N_f \int_{\vec{p},Q} Q^a
 \Bigl[n_{f}(\omega_p + g A_0 - \mu) -
       n_{f}(\omega_p - g A_0 + \mu) \Bigr], \\
 j^a_i & = & 0.
\ea
Solving now \eq\nr{SM} and writing
$\delta S_M^{f} =  \int_x \delta {\cal L}_M^{f}$, we obtain
\ba
 \delta {\cal L}_M^{f}  & = & 2 N_f T \int_{\vec{p},Q}
 \biggl[
 \ln\biggl(1+e^{({-\omega_p- gA_0+\mu})/{T}} \biggr)+
 \ln\biggl(1+e^{({-\omega_p+ gA_0-\mu})/{T}} \biggr)
 \biggr]. \la{dLf}
\ea
The integral over $\vec{p}$ is easily carried out,
and we finally arrive at
\ba
 \delta {\cal L}_{M}^{f}  &  = &  {N_f} \int_Q \biggl[
 \biggl( \fr72 \frac{\pi^2}{90} T^4 + \fr16 \mu^2 T^2+
 \frac{1}{12\pi^2}\mu^4 \biggr)
 \nn
 & - & g\frac{\mu}{3} \Bigl(T^2 + \frac{\mu^2}{\pi^2}\Bigr)
 A_0 + \frac{g^2}{2} \Bigl(\frac{T^2}{3} + \frac{\mu^2}{\pi^2}\Bigr)
 A_0^2 - \mu \frac{g^3}{3\pi^2} A_0^3
 + \frac{g^4}{12\pi^2} A_0^4 \biggr]. \la{LMf}
\ea
We have for completeness kept here even the field independent part,
accounting for the leading 1-loop (``free'') expression for the
fermionic contribution to the pressure of QCD.

Going to Euclidean metric by writing $A_0^M = i A_0^E$,
${\cal L}_E = -{\cal L}_M (A_0^M \to i A_0^E)$, we observe
immediately that \eq\nr{LMf} {\em would} agree with the
fermionic contribution to the dimensionally
reduced effective action~\cite{mu} for $A_0$;
or, equivalently, with the full 1-loop fermionic
contribution to the effective potential
for the phase of the Polyakov line~\cite{cka};
{\em provided} that
\be
 \int_Q Q^{a_1} Q^{a_2} \ldots Q^{a_n} \equiv
 (-1)^n \Bigl[\tr T^{a_1} T^{a_2} \ldots T^{a_n}
 \Bigr]_\rmi{symmetric part}, \quad
 n = 0, ..., 4.
\la{traces}
\ee
(The trivial factor $(-1)^n$ could be removed
by inverting the sign convention for $\mu$ here, or in~\cite{mu}.)
Thus the question is whether \eq\nr{traces} is satisfied. It is easy 
to show that it is for $n \leq 3$, while in general it is not for $n=4$.

Indeed, since the integration measure is gauge
invariant~\cite{kelly} and the $Q^a$'s commute, the result of the
left-hand-side of \eq\nr{traces} must have a covariant structure.
For SU(2) this is of the form 
\be 
\int_Q Q^{a_1} Q^{a_2} Q^{a_3} Q^{a_4} = 
 L(R) (\delta^{a_1 a_2} \delta^{a_3 a_4} + \delta^{a_1 a_3} 
 \delta^{a_2 a_4} + \delta^{a_1 a_4} \delta^{a_2 a_3}) \ . 
\ee
The integral here can be carried out even explicitly, for SU(2).
Alternatively, to fix the constant $L(R)$, we can contract this equation
with $\delta^{a_1 a_2} \delta^{a_3 a_4}$, and sum over indices.
Then the integral can easily be performed, due to the constraint
$\delta(Q^aQ^a - C_2(R))$ where $C_2(R)$ is the quadratic Casimir,
and the normalisation $\int_Q =d_R$ where $d_R$ is the dimension
of the representation. One finds $L(R) =({d_R}/{15}) C_2^2 (R)$.
Therefore, in the classical effective action the piece quartic in
$A_0$ reads 
\be 
 \int_Q A_0^4 = \frac{1}{5} d_R\, C_2^2(R)
 \left(A_0^a A_0^a \right)^2  , \label{class-tr} 
\ee 
while the
quantum result, for an arbitrary representation of SU(2), can be
seen to be 
\be 
 \tr A_0^4 = \frac{1}{5} d_R \, C_2(R) \left(C_2(R)
-\frac{1}{3} \right) \left(A_0^a A_0^a \right)^2 . \label{quan-tr}
\ee 
Writing $d_R = 2 j +1$, $C_2(R) = j \left(j+1 \right)$, we see
that only for high values of $j$ is the classical result
in~\eq\nr{class-tr} a good approximation of the quantum one.
(Suggestions along these lines were made already in~\cite{Wong,uh1}.) 
While the expressions given here hold for $N_c=2$, we expect similar
conclusions for any SU($N_c$), whereas for U(1) there is a perfect match.

The reason why operators below the quartic one are correctly
reproduced, is that the integral $\int_Q(...)$ involves
explicitly the quadratic and cubic Casimirs, which fix
the symmetric parts of the traces of two or three SU($N_c$) generators.
The quartic term, on the other hand, depends on the
commutation/anticommutation relations, and it is then at this order
that one sees a difference between the quantum and classical colour algebras.

While we have here focused our attention on the
``first'' formulation of classical kinetic theory,
a similar conclusion can be reached for the ``second''
formulation, based on~\eqs\nr{jmua1}, \nr{jmua}, although
the numerical discrepancy we find is different. In this 
case we do not know of an exact solution, but solve the 
equations iteratively, as was done to second order in~\cite{bl}.
At the third order, relevant for $\tr A_0^4$, we no longer
find a solution local in space, unless we take $A_0$ completely constant. 
Then, the result is too large by a factor three, for SU(2).

\section{Conclusions}

We have recalled in this brief note
that for some special backgrounds, like a
static one, the Boltzmann equation in~\eq\nr{boltzmann} can be
solved exactly, and the corresponding gauge field effective action
can be computed. We have then compared the result with the 1-loop
dimensionally reduced effective action for quantum field theory.
We remark that while the quadratic and cubic terms are correctly reproduced,
the classical non-Abelian
colour algebra generically fails at the next level. 
The relative discrepancy is the smaller the higher the dimensionality
of the colour representation; however, for the physical QCD case, 
some more elaborate formulation seems to be required in order to have
an exact match also at this level (see, e.g.,~\cite{egv}).

Of course, all this does not imply that classical kinetic 
theory could not be a useful tool for a non-perturbative
determination of many important observables of QCD, 
such as plasmon frequencies, damping rates, or physics 
related to the colour conductivity, for which effects from
the quartic coupling are subdominant. Nevertheless, our 
observation should underline the need for a better
beyond-the-leading-order understanding of the proper effective
description of real-time dynamics in the QCD plasma, which would
ideally at the same time also allow for a proper fixing of the
renormalisation scale in the gauge coupling $g$
as in the dimensionally reduced
theory~\cite{hl}, as well as for a systematic approach to the
continuum limit~\cite{db2}.


\section*{Acknowledgements}

We thank M. Tytgat for a very enlightening comment, and J.L.F.~Barbon, 
D. B\"odeker and M. Garcia-Perez for discussions. 
M.L.\ was partly supported by the TMR network {\em Finite
Tempe\-ra\-ture}  {\em Phase Transitions in Particle Physics}, EU Contract
No.\ FMRX-CT97-0122, and C.M. was supported by the EU
through the Marie-Curie Fellowship HPMF-CT-1999-00391.


\end{document}